\documentstyle[aps,twocolumn,epsfig]{revtex}
\begin{document}
\newcommand{\ve}[1]{\mbox{\boldmath $#1$}}
\twocolumn[\hsize\textwidth\columnwidth\hsize
\csname@twocolumnfalse%
\endcsname

\draft

\title {Damping of Hydrodynamic Modes in a Trapped Bose Gas
above the Bose-Einstein Transition Temperature}
\author{G. M. Kavoulakis$^1$, C. J. Pethick$^{1,2}$, and H. Smith$^3$}
 
\address{$^1$Nordita, Blegdamsvej 17, DK-2100 Copenhagen \O, Denmark, \\
        $^2$Department of Physics, University of Illinois at
        Urbana-Champaign, 1110 W. Green Street, Urbana, IL 61801-3080, \\
        $^3$\O rsted Laboratory, H. C. \O rsted Institute,
         Universitetsparken 5, DK-2100 Copenhagen \O, Denmark} 
\maketitle

\begin{abstract} 

    We calculate the damping of low-lying collective modes of a trapped 
Bose gas in the hydrodynamic regime, and show that this comes solely 
from the shear viscosity, since the contributions from bulk viscosity 
and thermal conduction vanish. The hydrodynamic expression for the 
damping diverges due to the failure of hydrodynamics in the outer 
parts of the cloud, and we take this into account by a physically 
motivated cutoff procedure.  Our analysis of available experimental data 
indicates that higher densities than have yet been achieved are necessary 
for investigating hydrodynamic modes above the Bose-Einstein transition 
temperature. 
\end{abstract}

\vskip2pc]

   In recent experiments on magnetically-trapped atomic vapors, alkali atoms
\cite{Anderson,Ketterle,Bradley} have been cooled to temperatures at which
they are degenerate and indeed Bose-Einstein condensation  has
been observed in them.   Frequencies and damping rates of collective modes 
in these systems have been investigated, both above and below the 
Bose-Einstein transition temperature, $T_c$ \cite{Kettosc,Cornellosc,C2}.  In 
this Letter we shall focus on properties above $T_c$. One can distinguish 
two regimes, the hydrodynamic one, for which the characteristic mode 
frequency is small compared with the collision frequency and the wavelength 
of the mode is large compared with the atomic mean free path, and 
the opposite limit, the collisionless one, for which collisions are 
relatively unimportant.  The frequencies of modes in the hydrodynamic 
regime have been calculated in Ref. \cite{Griffinosc}, and here we 
calculate their damping.   

    We begin by giving a simple derivation of the basic hydrodynamic 
equations.  Our treatment is essentially that of Ref. \cite{Landau} 
generalized to take into account the potential of the trap, and, since we 
are interested in small oscillations, we shall consider the linearized 
equations.   The Euler equation for the fluid velocity 
${\bf v}({\bf r}, t)$ is
\begin{eqnarray}
	m n_0({\bf r}) \frac {\partial {\bf v}} {\partial t} = 
   - {\ve \nabla} p({\bf r},t) + m n({\bf r},t) {\bf f},
\label{newton}
\end{eqnarray}
where $m$ is the mass of the atoms, $n({\bf r},t)$ is the particle density,
$n_0({\bf r})$ is the equilibrium particle density,
$p({\bf r},t)$ is the pressure and ${\bf f}$ is the force per unit
mass due to the external potential $U_0(\bf r)$, ${\bf f} = - {\ve \nabla} 
U_0({\bf r})/m$. In equilibrium, where the pressure is $p_0({\bf r})$, 
Eq. (\ref{newton}) implies that ${\ve \nabla} p_0({\bf r}) = m 
n_0({\bf r}) {\bf f}$. Taking the time-derivative of Eq. (\ref {newton})
and using the continuity equation, one finds
\begin{eqnarray}
        m n_0({\bf r}) \frac {\partial^2 {\bf v}} {{\partial t}^2} = 
     - {\ve \nabla} \frac {\partial p({\bf r},t)} {\partial t}
    - {\ve \nabla} \cdot [n_0{(\bf r}) {\bf v}] \, m {\bf f}. 
\label{newton2}
\end{eqnarray}
We calculate the first term on the right hand side
of Eq. (\ref{newton2}) by using the energy conservation 
condition \cite{Landau},
\begin{eqnarray}
        \frac {\partial} {\partial t} (\rho \epsilon) 
     = - {\ve \nabla} \cdot (w \rho {\bf v}) +
       \rho {\bf v} \cdot {\bf f},
\label{enecons}
\end{eqnarray}
where $\rho$ is the mass density and $\epsilon$ and $w$ are, respectively,
the internal energy and the enthalpy of the fluid per unit mass. 
Since we assume local thermodynamic equilibrium and neglect contributions 
to the energy due to interparticle interactions, we may use the results 
$p = \rho (w - \epsilon)$ and $\rho \epsilon = 3p/2$, and find that
\begin{eqnarray}
	\frac {\partial p({\bf r},t)} {\partial t} = 
   - \frac 5 3 {\ve \nabla} \cdot [p_0({\bf r}) {\bf v}] + 
  \frac 2 3 n_0({\bf r}) {\bf v} \cdot m {\bf f}.
\label{pressure}
\end{eqnarray}
Combining Eqs. (\ref{newton2}) and (\ref{pressure}), 
and using the fact that ${\ve \nabla} n_0({\bf r})$ is proportional to
${\ve \nabla} U_0({\bf r})$, we obtain for the equation of motion for 
${\bf v}({\bf r},t)$,
\begin{eqnarray}
          {\partial^2  {\bf v}\over \partial t^2} &=&
       {5\over 3}{p_0({\bf r}) \over {mn_0({\bf r})}} {\ve \nabla}
     [{\ve \nabla}\cdot {\bf v}]
   +{\ve \nabla}[{\bf v}\cdot
   {\bf f}  ]
+ {2\over 3}[{\ve \nabla}\cdot{\bf v}]
    {\bf f}  .
\label{velocity}
\end{eqnarray}
Equation (\ref{velocity}) has previously been derived by Griffin {\it et 
al.} \cite{Griffinosc} using kinetic theory. 

     In our present discussion we assume that the potential is axially 
symmetric,
\begin{eqnarray}
        U_0({\bf r}) = \frac 1 2 m \omega_0^2 (x^2 + y^2 + \lambda z^2),
\label{potential}
\end{eqnarray}
where $\omega_0$ is the frequency of the trap in the $x-y$ plane;
$\lambda \equiv \omega_z^2/\omega_0^2$ -- where $\omega_z$ is the 
frequency along the $z$-axis -- expresses the anisotropy of the trap.
Most experiments on oscillations have been done in such traps, but
our results may be generalized to the case of traps with no axis of symmetry.
As shown in Ref. \cite{Griffinosc}, in the lowest modes, which in a 
spherical trap correspond to monopole and quadrupole vibrations, the 
velocity field has the form ${\bf v}({\bf r},t) = {\bf v}({\bf r}) \cos(\omega
t)$, where ${\bf v}({\bf r}) = a(x \hat{\bf{x}} + 
y \hat{\bf{y}}) + b z \hat{\bf{z}}$. Here $a$ and $b$ are constants, and 
the frequencies, $\omega$, of the two modes are 
\begin{eqnarray}
    \left( \frac \omega {\omega_0} \right)^2 =
  \frac 1 3 \left[ 4 \lambda + 5 \pm (16 \lambda^2 - 
 32 \lambda + 25)^{1/2} \right],
\label{freq}
\end{eqnarray}
and $b / a =  3\omega^2 /2  \omega_0^2  - 5$.
  
   We turn now to the damping of these modes.  We adopt the standard approach 
of evaluating the rate of change of the mechanical energy, $E_{\rm mech}$, 
associated with the mode, which is given by \cite{Ldamp}
\begin{eqnarray}
          \dot {E}_{\rm mech} =  - \int\frac {\kappa} T 
          |\nabla T|^2 d {\bf r} - \int \zeta
({\ve \nabla} \cdot {\bf v})^2 d {\bf r}  
\nonumber \\ 
     - \int \frac {\eta} 2 
       \left( \frac {\partial v_i} {\partial x_k} + 
      \frac {\partial v_k} {\partial x_i} - \frac 2 3 \delta_{i,k}
   {\ve \nabla} \cdot  
    {\bf v} \right)^2  d {\bf r},
\label{energydot}
\end{eqnarray}
where $\kappa$ is the thermal conductivity,  $\eta$ is the  first, or shear, 
viscosity, $\zeta$ is the second, or bulk, viscosity and 
$T$ is the temperature. Because the system is inhomogeneous, 
the transport coefficients are generally spatially dependent.
Next we show that the shear viscosity is the only source of damping 
of the low-lying modes described above. The contribution from thermal
conduction vanishes because there are no temperature gradients for these modes, 
and the contribution from the bulk viscosity vanishes because $\zeta$ vanishes. 
To demonstrate the absence of temperature gradients, we observe that for the 
modes under consideration, which have the velocity field given above, ${\ve 
\nabla} \cdot {\bf v}=(2a+b) \cos(\omega t)$ is independent of 
position. Thus, from the continuity  
equation, $ dn/dt  + n {\ve \nabla} \cdot {\bf v} = 0$,
it follows that $(dn/dt)/ n$ is also constant. For an adiabatic
process in a free monatomic gas, $n \propto T^{3/2}$, and therefore the
deviation of the temperature from its equilibrium value is spatially 
independent. Consequently there are no temperature
gradients generated by the mode, and hence
no dissipation due to thermal conduction.  As to the second viscosity, this 
vanishes because a spatially-homogeneous, non-relativistic monatomic gas in 
equilibrium subjected to a slow uniform dilation (${\bf r}\rightarrow 
\nu{\bf r}$) remains in equilibrium, but at a different temperature.  
This result is independent of the statistics of the atoms and of the degree of 
degeneracy, and it is discussed in Ref. \cite{Henrik}. Inserting into 
Eq. (\ref{energydot}) the expression for the velocity field, we arrive at the 
following simple expression for the time-average of the rate of loss of 
mechanical energy:
\begin{eqnarray}
   \langle {\dot {E}}_{\rm mech} \rangle =
     - \frac 2 3 (a - b)^2 \int \eta({\bf r}) d {\bf r}.
\label{energydotexact}
\end{eqnarray}
Equation (\ref{energydotexact}) implies that 
$\langle {\dot {E}}_{\rm mech} \rangle$
vanishes for the monopole mode for the isotropic case ($\lambda =1$).

    The next task is to calculate the first viscosity. At the low energies of 
interest in experiments, the scattering of two atoms is purely $s$-wave, and 
the total cross section is $\sigma = 8\pi a_{\rm scat}^2$, where $a_{\rm scat}$ 
is the scattering length.  We consider the viscosity in the classical 
limit, since we expect the classical limit to be quantitatively accurate,
even at $T$ very close to $T_c$, from comparison to the known effects of
degeneracy on the heat capacity. The viscosity has the general form 
\cite{Henrik}
\begin{eqnarray}
  \eta =C_{\eta} \frac {(mkT)^{1/2}} {\sigma}.
\label{dd}
\end{eqnarray}
A simple relaxation-time approximation, with a scattering rate equal to
$n_0({\bf r}) \sigma v$, where $v = (2 \epsilon/m)^{1/2}$
is the particle velocity and $\epsilon$ is the single-particle energy,
leads to the result $C_\eta = 2^{7/2}/(15\pi^{1/2})\approx 0.426$, while 
a variational calculation gives $C_\eta=5\pi^{1/2}/2^4 \approx 0.554$.
 
An important feature of the expression for the viscosity is its independence of 
the particle density.  Consequently, the integral in Eq. (\ref{energydotexact})
formally diverges at large distances from the center of the trap. The reason    
for this is that hydrodynamics fails, because in the outer parts of the cloud, 
particle mean free paths are too long for hydrodynamics to be applicable.  To 
solve this problem, one should treat the outer parts of the cloud using kinetic 
theory, rather than hydrodynamics, but to obtain a first estimate 
of the effects we shall assume
that the hydrodynamic description holds out to a distance such 
that an atom incident from outside the cloud has a probability of no more than 
$1/e$ of not suffering a collision with another atom. In the parlance of 
radiative transport, this corresponds to an optical depth of unity.  
Mathematically, this condition is
\begin{eqnarray}
   1 \approx \int_{r_{0,s}}^{\infty} \frac {ds} {l({\bf r})},
\label{r0}
\end{eqnarray}
where the local mean free path is $l({\bf r}) = [n_0({\bf r}) \sigma]^{-1}$,
and $r_{0,s}$ is the cutoff, which depends on direction. The integral
in Eq. (\ref{r0}) is to be performed along the path for which the density
gradient is steepest, that is along ${\ve \nabla} U_0$, and $ds$ is the
corresponding line element. For a classical distribution, $n_0({\bf r}) = n(0)
e^{-U_0({\bf r})/ k_B T}$, where $n(0)$ is the density at the center.
Since $d U_0 = |{\ve \nabla} U_0| ds$, Eq. (\ref{r0}) can be written as
\begin{eqnarray}
   1 \approx n(0) \sigma k_B T \frac 1 {|{\ve \nabla} U_0({\bf r}_{0,s})|}
  e^{-U_0({\bf r}_{0,s})/ k_B T}.
\label{r00}
\end{eqnarray}
For large values of the dimensionless parameter 
$\lambda^{1/2} N a_{\rm scat}^2 m \omega_0^2/ k_B T$, Eq. (\ref{r00}) gives
\begin{eqnarray}
   r_{0,s}^2 \approx 2 \frac {k_B T} {m \omega_0^2} \,
    \frac 1 {\sin^2 \theta + \lambda \cos^2 \theta} \ln \tau_s(\theta),
\label{r0approx}
\end{eqnarray}
where $\theta$ is the angle of ${\bf r}_{0,s}$ with respect to the 
$z$-axis. The dimensionless quantity $\tau_s(\theta)$ is essentially
the total optical depth at the center of the cloud  
\begin{eqnarray}
    \tau_s(\theta) \equiv \tau_0 \lambda^{1/2}
         \left( \frac {\sin^2 \theta + \lambda \cos^2 \theta}
            {\sin^2 \theta + \lambda^2 \cos^2 \theta} \right)^{1/2},
\label{def}
\end{eqnarray}
where
\begin{eqnarray}
   \tau_0 = \sigma n(0) \left( \frac {k_B T} 
  {2 m \lambda \omega_0^2} \right)^{1/2}.
\label{aa}
\end{eqnarray}
The density at the center of the cloud is $n(0) = N {\lambda}^{1/2}
(m \omega_0^2/2 \pi k_B T)^{3/2}$. 
The volume of the atomic cloud is given by integrating $r_{0,s}^3/3$, 
Eq. (\ref{r0approx}), over the solid angle. Equation (\ref{energydotexact})
with the variational estimate for the viscosity ($C_{\eta} = 5 \pi^{1/2}/2^4$)
can then be written as 
\begin{eqnarray}
    \langle {\dot {E}}_{\rm mech} \rangle \approx
  - \frac {5 \pi^{1/2}} {2^{7/2} 3^2} (a - b)^2
    \frac {(k_B T)^2} {m \omega_0^3 a_{\rm scat}^2} 
  f(\lambda, \tau_0).
\label{energydott}
\end{eqnarray}
The function $f(\lambda, \tau_0)$ is defined as
\begin{eqnarray}
   f(\lambda, \tau_0) &=& \int_{-1}^1
  \frac  {dx} {[(\lambda -1)x^2 + 1]^{3/2}}
\times \nonumber \\ &\times&
 \left[ \ln \left( \tau_0 \lambda^{1/2} \left( \frac {(\lambda -1)x^2 + 1}
  {(\lambda^2 -1)x^2 + 1} \right)^{1/2} \right) \right]^{3/2}.
\label{fun}
\end{eqnarray}
We expect the leading term $(\propto [\ln \tau_0]^{3/2})$ to be asymptotically
exact for large $\tau_0$, but the value of the cutoff in the logarithm
will depend on the detailed kinetic-theory solution to the boundary-layer
problem.

\begin{figure}
\begin{center}
\epsfig{file=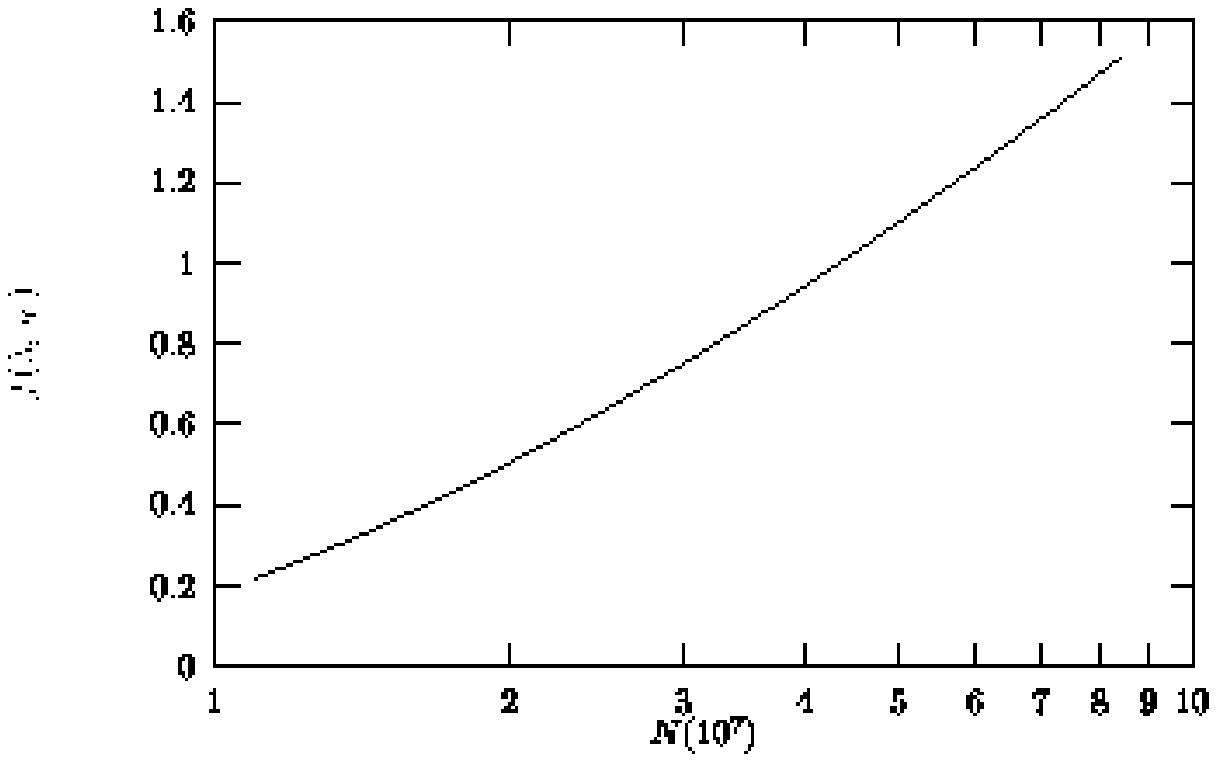,width=\linewidth}
\begin{caption}
{We plot $f(\lambda, \tau_0)$ from Eq. (\ref{fun})
as function of the number of particles $N$,
for $\lambda = 8$, $a_{\rm scat} = 53$ \AA,
$\nu_0 = 129$ Hz, and $T=1.3 \,T_c$, which are the parameters
in the experiment of Ref. [6].
Asymptotically, for $\tau_0 \gg 1$,
$f(\lambda=8, \tau_0) \approx 2
\lambda^{-1/2} [\ln (\tau_0 \lambda^{1/2})]^{3/2}
- 0.53 \, [\ln (\tau_0 \lambda^{1/2})]^{1/2}$.}
\end{caption}
\end{center}
\label{FIG1}
\end{figure}

   To calculate the damping rate we need to evaluate the mechanical energy of 
the cloud, $E_{\rm mech}$. In any oscillator the time-average
of $E_{\rm mech}$ is equal to the maximum kinetic energy in the mode, so
\begin{eqnarray}
   \langle {E}_{\rm mech} \rangle = \frac 1 2 \int m n_0({\bf r}) v^2({\bf r})
d{\bf r},
\label{bb}
\end{eqnarray}
and therefore,
\begin{eqnarray}
   \langle {E}_{\rm mech} \rangle = N \frac {k_B T} {\omega_0^2}
   \left( a^2 + \frac {b^2} {2 \lambda} \right).
\label{finenen}
\end{eqnarray}

   We can now introduce the {\it amplitude} damping rate $\tau_{\rm damp}^{-1}$
(as opposed to the {\it energy} damping rate, which is twice the amplitude
damping rate), given by the absolute value of the ratio
$\langle {\dot {E}}_{\rm mech} \rangle / 2 \langle {E}_{\rm mech} \rangle$,
\begin{eqnarray}
     \tau_{\rm damp}^{-1} \approx
    \frac {5 \pi^{1/2}} {2^{9/2} 3^2}
   \frac {k_B T} {m \omega_0 a^2_{\rm scat}}
  \frac 1 N
 \frac {(a-b)^2} {\left( a^2 + b^2/2 \lambda \right)} f(\lambda, \tau_0).
\label{tau0}
\end{eqnarray}
In order to elucidate the origin of the damping of the modes given by
Eq. (\ref{freq}), we remark that when the viscosity is taken to be constant,
the non-equilibrium stress tensor is constant in space,
and therefore does not give rise to any force on a fluid element.
The damping is caused by the entropy generated by the non-equilibrium
energy current density, which has a constant, non-zero divergence.

   Let us now turn to the experiments on sound propagation that have been
performed to date. A necessary condition for hydrodynamics to be applicable
is that the mean free path be small compared with the characteristic length
scale of the mode. For the low modes we are studying here, this is equivalent
to the requirement that the optical depth $\tau_s(\theta)$ be large compared
with unity in all directions. For the MIT experiment \cite{Kettosc}, for
which $\lambda < 1$, this condition implies that $\lambda^{1/2} \tau_0 \gg 1$.
In terms of the characteristic lengths, $R_z = (2 k_B T/m \omega_z^2)^{1/2}$ 
and $R_{\perp} = (2 k_B T/m \omega_0^2)^{1/2}$, which measure the 
spatial extent of the cloud along the $z$-axis and perpendicular to 
it, the condition $\lambda^{1/2} \tau_0 \gg 1$ is seen to be equivalent to the
requirement $R_{\perp} \gg l(0)$, where $l(0)$ is the mean free path in the
center of the cloud. For the parameters of this experiment \cite{Kettosc},
the minimum number of particles that is required for this condition to
be satisfied is $\approx 3 \times 10^{9}$, whereas the number 
of particles at $T = 2 \, T_c$ is found to be $N \approx 5 \times 10^7$
\cite{private}. For the parameters in the JILA experiment \cite{Cornellosc,C2},
the minimum value of $\tau_s(\theta)$ is $\tau_0$, since $\lambda > 1$.
In this case the condition $\tau_0 \gg 1$ is equivalent to $R_z \gg l(0)$.
Therefore, for the JILA experiment, the minimum
number of particles required to attain hydrodynamic conditions
is $\approx 1 \times 10^7$ at $T=1.3 \,T_c$, which is to be compared
with the experimental value $N \approx 8 \times 10^4$ at $T \approx 1.3
\, T_c$ \cite{C2}.  Figure 1 shows $f(\lambda, \tau_0)$
as function of $N$, with all the other parameters equal to the ones of
Ref. \cite{C2} and $T=1.3\,T_c$. The lowest value of $N$ is chosen 
to be $N = 1.1 \times 10^7$, i.e., the one for which $\tau_0 = 1$.   

    Now we estimate for the MIT and JILA experiments the magnitude of the 
characteristic lengths $R_z$ and $R_{\perp}$, and compare them with the mean
free path in the center of the cloud. The transition temperature is
obtained from $k_B T_c = 0.94 \, N^{1/3} \hbar (\omega_0^2 \omega_z)^{1/3}$. 
For the MIT experiment $N \approx 2.5 \times 10^7$ at $T_c$ \cite{private},
while $\omega_0 \approx 2 \pi \times 250$ Hz and $\lambda \approx 5.8
\times 10^{-3}$ \cite{Kettosc}. At $T = 2 \, T_c$ the particle number is
$\approx 5 \times 10^7$, resulting in $R_z \approx 380$ $\mu$m and 
$R_\perp \approx 29$ $\mu$m. The corresponding value of the central density
$n(0) = N / (\pi^{3/2} R_z R_{\perp}^2)$ is $\approx 2.9 \times 10^{13}$ 
cm$^{-3}$. With the scattering cross section $\sigma \approx
1.9 \times 10^{-12}$ cm$^2$ ($a_{\rm scat} \approx 28$ \AA) \cite{Dav}, 
we obtain $l(0) = [n(0) \sigma]^{-1} \approx 180$ $\mu$m, which is
less than $R_z$, but much larger than $R_{\perp}$. The other condition
for hydrodynamic behaviour is that the frequency of the mode be small
compared with an average particle scattering rate. A single particle
undergoes collisions at a rate $n_0({\bf r}) \sigma v_{\rm th}$, where 
$v_{\rm th} = (8 k_B T/\pi m)^{1/2}$ is the average particle velocity.
The average scattering rate in the cloud is thus $\tau_{\rm scat}^{-1} = \int
n_0^2({\bf r}) \sigma v_{\rm th} \, d{\bf r} / \int n_0({\bf r}) \, d{\bf r}
= n(0) \sigma (k_B T/\pi m)^{1/2}$, since $\int n_0^2({\bf r}) \, d{\bf r} /
\int n_0({\bf r}) \, d{\bf r} = n(0)/2^{3/2}$. For the MIT experiment we 
therefore estimate $\omega \tau_{\rm scat} \approx 2.2$ at $T = 2 \, T_c$.

    For the JILA experiment $\omega_0 \approx 2 \pi \times 129$ Hz,
$\lambda = 8$ and the particle number at $T_c$ is $N \approx 4 \times
10^4$ \cite{C2}. At $T = 1.3 \, T_c$, where $N \approx 8 \times 10^4$, we 
obtain $R_z \approx 4$ $\mu$m and $R_{\perp} \approx 10$ $\mu$m, while the
central density is $n(0) \approx 3.7 \times 10^{13}$ cm$^{-3}$. The
mean free path $l(0)$ is estimated to be 38 $\mu$m corresponding to a 
scattering cross section $\sigma \approx 7 \times 10^{-12}$ cm$^2$.
Finally the dimensionless parameter $\omega \tau_{\rm scat}$ is $\approx 19$.
We should mention that using Eq. (\ref{tau0}) we find that for $N = 1.1 
\times 10^7$, the amplitude damping times for the JILA experiment are
$\approx 8$ ms and 47 ms at $T = 1.3 \, T_c$ for the modes which
correspond to monopole and quadrupole vibrations in a spherical trap, 
respectively.

   We thus conclude that even though in the MIT experiment $l(0) 
{\raisebox{-.5ex}{$\stackrel{<}{\sim}$}} R_z$,
there are as yet no experiments with which we can directly compare our
calculation of the damping. To obtain a semi-quantitative description, we
adopt a phenomenological interpolation formula for the
frequency and the damping rate of the modes. This has the usual form,
\begin{eqnarray}
     \omega^2 = \omega_C^2 + \frac {\omega_H^2 - \omega_C^2}
   {1 - i \omega \tau},
\label{interpolation}
\end{eqnarray}
characteristic of relaxation processes. Here
$\omega_C$ is the frequency of the mode in the collisionless
regime and $\omega_H$ in the hydrodynamic regime, and $\tau$ is 
a characteristic relaxation time, which we anticipate will be of order
the scattering time. Equation (\ref{interpolation})
gives the qualitatively correct limiting behaviour
for $\omega \tau \gg 1$ and $\omega \tau \ll 1$. 
The imaginary part of the above equation gives for the damping rate
\begin{eqnarray}
   \tau_{\rm damp}^{-1} = \frac 1 2 
  \frac {(\omega_C^2 - \omega_H^2) \tau} 
  {1 + (\omega \tau)^2}.
\label{phdamp}
\end{eqnarray}

    The amplitude damping time in the MIT experiment \cite{Kettosc}
is measured to be $\approx 80$ ms at $T = 2 \,T_c$.
The frequency $\nu_C = 2 \nu_z \approx 38$ Hz; Eq. (\ref{freq})
gives for $\nu_H \approx (12/5)^{1/2} \nu_z \approx 30$ Hz.
Using these numbers and Eq. (\ref{phdamp}), we can solve for 
$\omega \tau$ to get the two solutions 
3.6 and $3.6^{-1} \approx 0.28$. We regard the larger
solution as being the physically relevant one, since it is close to
our estimate of $\omega \tau_{\rm scat} \approx 2.2$. The real part of
Eq. (\ref{interpolation}), with $\omega \tau = 3.6$, implies that
the oscillation frequency is $\approx 37.5$ Hz, which is consistent with 
the experimental value of $35 \pm 4$ Hz. For the JILA experiment
\cite{C2}, the amplitude damping time is $\approx 50$ ms at $T \approx 1.3 
\,T_c$. Since $\nu_C \approx 258$ Hz and $\nu_H \approx 221$ Hz, 
Eq. (\ref{phdamp}) implies that $\omega \tau \approx 10.6$ or $10.6^{-1}$.
Again we consider the larger value to be the physical one, since it is
close to our theoretical estimate above, $\omega \tau_{\rm scat} \approx 19$.
We thus conclude that damping of the modes in clouds of bosons above 
$T_c$ is in good agreement with theoretical expectations.

     Helpful discussions with Gordon Baym, W. Ketterle and D. M. Kurn
are gratefully acknowledged. G.M.K. would like to thank the Foundation
of Research and Technology, Hellas (FORTH) for its hospitality.

\end{document}